\documentclass[fleqn,usenatbib]{mnras}

\usepackage{newtxtext,newtxmath}
\usepackage[T1]{fontenc}

\DeclareRobustCommand{\VAN}[3]{#2}
\let\VANthebibliography\thebibliography
\def\thebibliography{\DeclareRobustCommand{\VAN}[3]{##3}\VANthebibliography}

\usepackage{graphicx}	% Including figure files
\usepackage{amsmath}	% Advanced maths commands
\usepackage{amssymb}	% Extra maths symbols

\usepackage{tabularx}

\usepackage{amsfonts}
\usepackage{graphicx}
\usepackage{subfigure}
\usepackage{mathtools}

\graphicspath{{./}{Figures_post_ref/}}

\title[CHEOPS \& TESS]{CHEOPS observations of TESS primary mission monotransits}

\author[Benjamin F. Cooke et al.]{
Benjamin F. Cooke,$^{1,2}$\thanks{E-mail: b.cooke@warwick.ac.uk}
Don Pollacco,$^{1,2}$
Monika Lendl,$^{3,4}$
\newauthor
Thibault Kuntzer$^{3}$
and Andrea Fortier$^{5}$
\\
$^{1}$Department of Physics, University of Warwick, Gibbet Hill Road, Coventry CV4 7AL, UK\\
$^{2}$Centre for Exoplanets and Habitability, University of Warwick, Gibbet Hill Road, Coventry CV4 7AL, UK\\
$^{3}$Observatoire de Gen{\`e}ve, Universit{\'e} de Gen{\`e}ve, 51 Ch. des Maillettes, 1290 Sauverny, Switzerland\\
$^{4}$Space Research Institute, Austrian Academy of Sciences, Schmiedlstr. 6, 8042 Graz, Austria\\
$^{5}$Centre for Space and Habitability (CSH), University of Bern, Gesellschaftstrasse 6, CH-3012, Bern, Switzerland
}

\date{Accepted XXX. Received YYY; in original form ZZZ}

\pubyear{2020}

\begin{document}
\label{firstpage}
\pagerange{\pageref{firstpage}--\pageref{lastpage}}
\maketitle

\begin{abstract}
We set out to look at the overlap between CHEOPS sky coverage and TESS primary mission monotransits to determine what fraction of TESS monotransits may be observed by CHEOPS. We carry out a simulation of TESS transits based on the stellar population in TICv8 in the primary TESS mission. We then select the monotransiting candidates and determine their CHEOPS observing potential. We find that TESS will discover approximately 433 monotransits during its primary mission. Using a baseline observing efficiency of 40\% we then find that 387 of these ($\sim$\,89\%) will be observable by CHEOPS with an average observing time of $\sim$\,60\,days per year. Based on the individual observing times and orbital periods of each system we predict that CHEOPS could observe additional transits for approximately 302 of the 433 TESS primary mission monotransits ($\sim$\,70\%). Given that CHEOPS will require some estimate of period before observing a target we estimate that up to 250 ($\sim$\,58\%) TESS primary mission monotransits could have solved periods prior to CHEOPS observations using a combination of photometry and spectroscopy.
\end{abstract}

\begin{keywords}
Planetary systems -- Surveys -- Planets and satellites: detection
\end{keywords}

%%%%%%%%%%%%%%%%%%%%%%%%%%%%%%%%%%%%%%%%%%%%%%%%%%

%%%%%%%%%%%%%%%%% BODY OF PAPER %%%%%%%%%%%%%%%%%%

\section{Introduction}
\label{sec:Introduction}

The CHaracterising ExOPlanets Satellite \citep[CHEOPS,][]{2013EPJWC..4703005B,2014SPIE.9143E..2JF} is an ESA mission dedicated to performing ultra-high precision photometry on known transiting planetary systems. CHEOPS was launched on 18$\rm^{th}$ December 2019 and has a nominal mission lifetime of 3.5 years \citep{2019SPIE11116E..05R}. CHEOPS will be able to very precisely measure radii for the planets it observes due to its photometric precision and short observing cadence (1 minute or better). Additionally, if observing systems with uncertain periods (or those with only bounding limits on the period), CHEOPS will be able to help confirm periods by capturing additional transits or ruling out period aliases.

These characteristics make CHEOPS an attractive tool for better characterising monotransiting systems discovered as part of the TESS primary mission. These are targets identified as transiting planets by the TESS primary mission but with only one observed transit. Without the ability to fold multiple transits these systems require significant effort to properly characterise but are of interest since they are generally have longer periods than other TESS detections (those that exhibit multiple transits) \citep{2018A&A...619A.175C}. Better characterisation of these systems is vital to help develop our understanding of how planetary characteristics vary with orbital separation \citep{2004ASPC..321..355N,2004ASPC..321..298J,2013A&A...558A.109A}. Moving further from the host additionally favours the detection of potential habitable zone planets, especially around solar type stars whose habitable zone periods are on the order of hundreds of days \citep{1993Icar..101..108K,2013ApJ...765..131K}.

The Transiting Exoplanet Survey Satellite \citep[TESS,][]{2015JATIS...1a4003R} is over a year into its primary mission having already moved into the northern ecliptic hemisphere. As part of its exoplanet yield TESS is expected to discover hundreds of monotransiting systems \citep{2018A&A...619A.175C,2019AJ....157...84V,2019A&A...631A..83C}. These systems will be known to host transiting exoplanets so will be valid targets for CHEOPS but will have no known period. These systems require time and effort to characterise but some systems have been successfully recovered with more on the way \citep{2019MNRAS.tmp.2805G,2019arXiv191005050L,2020arXiv200209311G}.% One of the main complications in determining the period for these systems is period alias \citep{2019A&A...631A..83C}. Even when a second transit is found the exact period may still not be known. Observations with CHEOPS will be able to better characterise these systems by reducing the number of period aliases, either by catching additional transits or confirming the absence of predicted transits.%. {\bf TK: how? by observing $\gtrsim$2 additional transits?}

CHEOPS will not be able to directly observe these systems based on a single TESS transit alone since this would require constant observation until the planet transits again. However, if a prior on the period was available, or a small number of period aliases were found, CHEOPS could instead target specific times and, using its high cadence and precision, vastly improve the transit, and thus planet, characteristics. To this end it is of interest to know what fraction of TESS primary mission monotransits CHEOPS could observe and for which of these period estimates could be found that would justify the use of CHEOPS time.

We set out our paper in the following way. Section \ref{sec:Methodology} describes the simulation population and the TESS and CHEOPS observations. Section \ref{sec:Results} shows the simulation results for both the TESS and CHEOPS observations. Section \ref{sec:Period estimation} shows our attempts to estimate periods for the TESS monotransits and Section \ref{sec:Conclusions} gives our conclusions.

\section{Methodology}
\label{sec:Methodology}

%The first step is to produce a set of TESS monotransit targets that can be tested for overlap with CHEOPS. We proceed as in \cite{2018A&A...619A.175C} and \cite{2019A&A...631A..83C}. 

\subsection{Simulation population}

%We set up our simulation using a similar method as discussed in \cite{Cooke2018}. We begin with our stellar catalogue, the TESS Input Catalogue (TIC) Candidate Target List (CTL) available from the Mikulski Archive for Space Telescopes (MAST\footnote{\href{http://archive.stsci.edu/tess/tic\_ctl.html}{http://archive.stsci.edu/tess/tic\_ctl.html}}). This simulation employs the latest version, version 8, of the CTL cross-matched against the TIC as described by \citet{Stassun2019}. We then filter this sample using TESS-band magnitude cuts, $3.0 \leq m_{TESS} \leq 17.0$, and effective temperature cuts, $2285 \leq T_{eff} \leq 10050$. Taking only those stars in the southern ecliptic hemisphere produces a final input stellar catalogue of 4789372 potential hosts.

%For each target we then predict how many TESS sectors will observe it. We simulate 13 sectors, each $24^\circ\times96^\circ$ on the sky, and, using ecliptic coordinates, find the number of sectors overlapping each star.

%With the stellar catalogue known we then simulate a planetary population around the stars in the same way as discussed in \cite{Cooke2018}. Occurrence rates as functions of radius and period are taken from \cite{Dressing2015} and \cite{Fressin2013} depending on spectral type and the hosts are randomly populated with planetary companions. Transit parameters such as depth and duration are calculated following the formalism set out in \cite{Cooke2018}, which use the equations from \cite{Winn2014} and \cite{Barclay2018}.

We produce our stellar and planetary populations as in \cite{2018A&A...619A.175C}. Our stellar populations we take as the TESS Input Catalogue (TIC) Candidate Target List (CTL) version 8 \citep{2019AJ....158..138S} available from the Mikulski Archive for Space Telescopes (MAST\footnote{\href{http://archive.stsci.edu/tess/tic\_ctl.html}{http://archive.stsci.edu/tess/tic\_ctl.html}}). The sample is filtered by magnitude in the TESS-band, $m_{TESS}$, and effective temperature, $T_{\rm eff}$, using $3.0 \leq m_{TESS} \leq 17.0$, and $2285 \leq T_{\rm eff} \leq 10050$\,K.

Planets are generated around these stars based on occurrence rates as functions of radius and period from \cite{2015ApJ...807...45D} (M-stars) and \cite{2013ApJ...766...81F} (AFGK-stars). We determine transit parameters as in \cite{2018A&A...619A.175C} using equations from \cite{2010arXiv1001.2010W} and \cite{2018ApJS..239....2B}.

\subsection{Detectability}

%S/N, and thus detectability, is a function of transit depth, stellar contamination, instrumental noise, and number of data points in transit. We calculate this value using the equations presented in \cite{Cooke2018}, employing the $5^{\rm th}$ order polynomial noise approximation given in \cite{Stassun2017}.

%Lastly, we determine host cadence using a priority metric,

%\begin{equation}
%\label{eq:metric}
%\frac{\sqrt{N_s}}{\sigma_{1hr} R_{\star}^{3/2}},
%\end{equation}

%where $N_s$ is the number of sectors for which a target is observed, $\sigma_{1hr}$ is the photometric noise in an hour and $R_{\star}$ is stellar radius. The top ranked 200,000 stars are given an observing cadence of 2\,min, corresponding to the Postage Stamp (PS) targets, with the rest getting 30\,min cadence during the first year of observing and 10\,min cadence when TESS returns to the south hemisphere in year 4 corresponding to the Full-Frame Images (FFIs).

%Although it is now known exactly which stars in the TIC have been observed at 2\,min cadence in Year 1 the metric used here should produce an almost identical population. Since we do not yet know the 2\,min targets for the extended mission it was decided to continue with the metric for both the primary and extended mission.

For TESS detectability we first determine how many TESS sectors will observe each TICv8 target. From there we use the transit parameters with a noise approximation to calculate a Signal-to-Noise ratio, $S/N$. We take our noise approximation from \cite{2018AJ....156..102S} who use a 5$\rm^{th}$ order polynomial fit. We take into account the arguments of \cite{2018ApJS..239....2B} and require $S/N \geq 10.0$ for a detectable observation. Finally we must determine which stars will be observed at 2\,min cadence and which will only receive 30\,min. We determine this using the priority metric

\begin{equation}
\label{eq:metric}
\frac{\sqrt{N_s}}{\sigma_{1\rm hr} R_{\star}^{3/2}},
\end{equation}

where $N_s$ is the number of sectors for which a target is observed, $\sigma_{1\rm hr}$ is the photometric noise in an hour and $R_{\star}$ is stellar radius. The 200,000 top priority targets receive 2\,min cadence observations.

As in \cite{2019A&A...631A..83C} we continue to use this metric as it is almost identical to the known 2\,min sample but remains unbiased into the northern ecliptic where the exact 2\,min targets are not yet known for all sectors.

\subsection{TESS observations}

To simulate TESS observations of our stellar population we use the same sector window functions method described in \cite{2018A&A...619A.175C}. This method involves taking the timestamps from actual TESS data for each available sector, stripping out data points with systematics and only simulating observations at the remaining times. This produces a more realistic observing strategy than was used in \cite{2018A&A...619A.175C}. The improvement between the method employed here and the \cite{2019A&A...631A..83C} implementation of this method is that data for more TESS sectors is available. We now have timestamps for sectors 1-15. We then extrapolate this data up to sector 26 by replicating the sector 15 timestamps separated by the average inter-sector gap to complete the primary mission. As an additional improvement we offset 6 northern sectors (sectors 14, 15, 16, 24, 25 and 26) to follow the procedures that TESS is employing to mitigate scattered light. Once the simulation is complete we select all those planets with one detectable transit during the TESS primary mission as our monotransit sample.

\subsection{CHEOPS observations}

\begin{figure}
    \begin{subfigure}[Minimum 19\,minutes per orbit ($\sim$\,20\%)
    \label{fig:coverage19}]
    {\includegraphics[width=0.85\columnwidth]{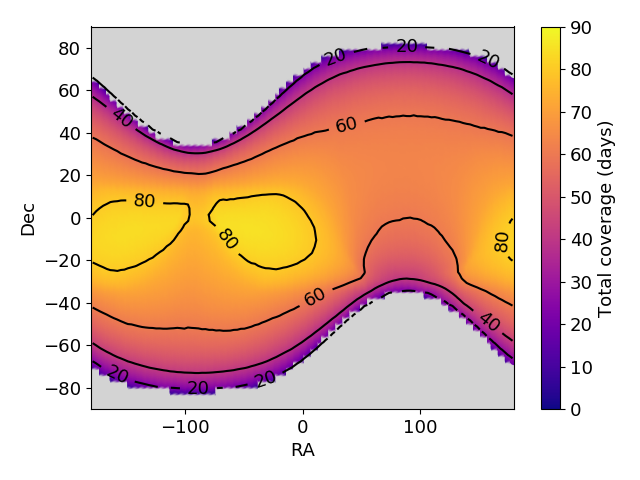}}
    \end{subfigure}
    \begin{subfigure}[Minimum 59\,minutes per orbit ($\sim$\,60\%)
    \label{fig:coverage59}]
    {\includegraphics[width=0.85\columnwidth]{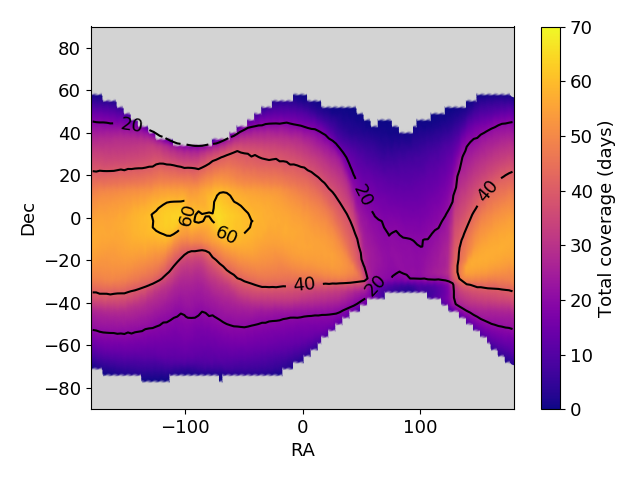}}
    \end{subfigure}
    \centering
    \begin{subfigure}[Minimum 98\,minutes per orbit ($\sim$\,99\%)
    \label{fig:coverage98}]
    {\includegraphics[width=0.85\columnwidth]{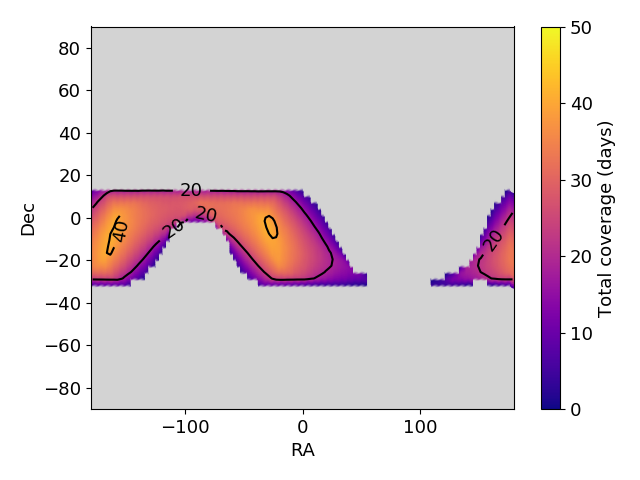}}
    \end{subfigure}
    \caption{CHEOPS sky coverage for three different values of minimum observing time per orbit. The colours denote total observing time in days (across one year) and the areas in grey are the regions that never reach the required coverage per orbit. Also shown are contours at 20, 40, 60 and 80\,days (where applicable).}
\label{fig:coverage}
\end{figure}

We then looked at whether these monotransits will be observable by CHEOPS. CHEOPS orbital period is 98.6\,minutes with an observing cadence of 1\,minute or better. The location of a target on the sky will lead to a different amount of interruption per orbit. Figure \ref{fig:coverage} shows three sky plots of CHEOPS coverage showing the total time that each part of the sky can be observed for in one year\footnote{Available from \href{https://www.cosmos.esa.int/web/cheops-guest-observers-programme/ao-1}{https://www.cosmos.esa.int/web/cheops-guest-observers-programme/ao-1}}. The difference between the three plots is the minimum amount of observing time required per orbit; 19, 59 and 98\,minutes (roughly corresponding to 20\%, 60\% and 99\% of the orbital period). The greater the required observing time per orbit, the smaller the total sky area that can be observed. However, the areas of sky that are observable can be observed for longer per orbit. This leads to longer observing times, but spread across a smaller region of sky.

The specific amount of observing time required per CHEOPS orbit is difficult to estimate due to unknown factors such as the specific variability of the host star, when in the orbital phase CHEOPS observations of a target will begin and what fraction of a transit may be caught. Because of these complications we have elected to require a baseline of 40\% % {\bf TK: why selecting 40\% rather than 50 or 30? Is there a quantifiable rationale behind it?}
observing efficiency for each TESS primary mission monotransit. That is, if a TESS monotransit falls on to a region of sky for which CHEOPS could observe with an efficiency of $\geq$\,40\% we consider that monotransit to be observable with CHEOPS. 40\% efficiency gives $\sim$40\,minutes of observation per orbit and was chosen as this matches well with the average length of an ingress/egress of the simulated monotransits. We do note that CHEOPS observability exhibits a slow continuous change between orbits (see Appendix C of \cite{2013arXiv1310.7800K}) but, providing we select targets allowing for this it should only increase the number of available targets. %{\bf TK: note that the observability will change from one orbit to another (it's a slow continuous change. See e.g. Appendix C of \href{https://arxiv.org/abs/1310.7800}{this document on the effect of stray light}).}

\section{Results}
\label{sec:Results}

\subsection{TESS}
\label{sec:TESS results}

\begin{figure}
    {\includegraphics[width=\columnwidth]{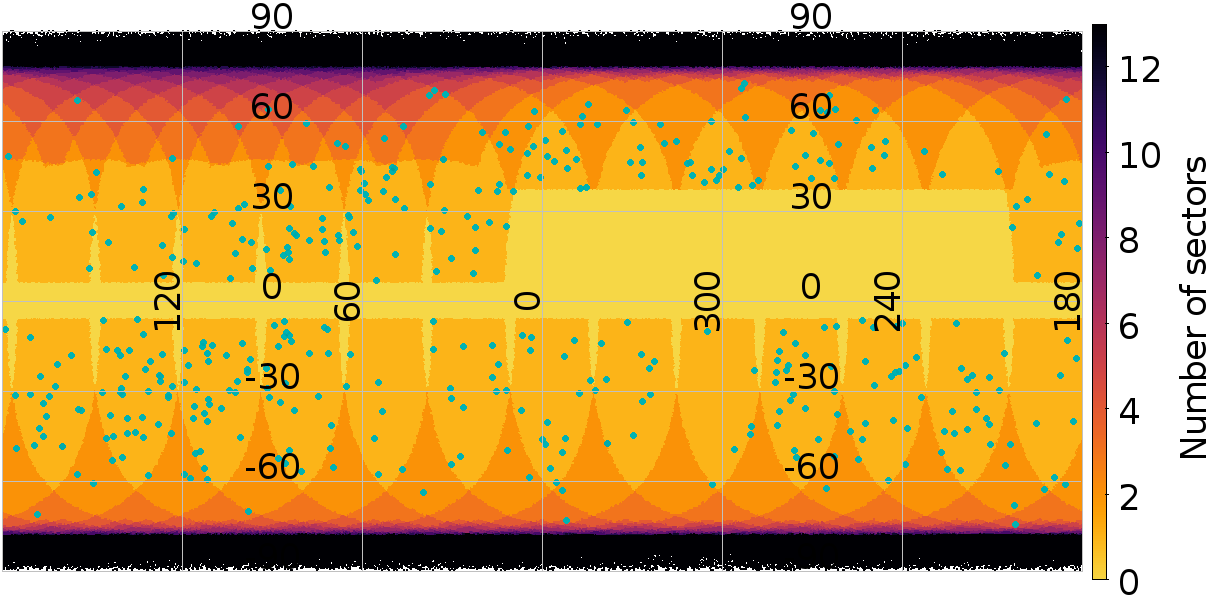}}
    \caption{All TICv8 stars coloured by number of TESS observing sectors. Monotransit hosts are overlaid in blue. The larger region with no coverage in the northern ecliptic is due to the sector offsets to combat scattered light.}
\label{fig:monos}
\end{figure}

The results of the TESS part of the simulation are shown here. From the TESS primary mission we predict 433 monotransits. That is, 433 of our simulated planets will be observed to transit once during TESS primary mission observations with an $S/N\geq10.0$. Figure \ref{fig:monos} shows a plot of all stars in TICv8 coloured by number of sectors with which TESS will observe them. The monotransiting planets are overlaid in blue. This value is comparable to the equivalent values presented in \cite{2018A&A...619A.175C} and \cite{2019A&A...631A..83C} allowing for the reduction in TESS coverage in the northern ecliptic hemisphere and an increased $S/N$.

% When we combine this data with the CHEOPS part of the simulation the first key result is to look at the distribution of ingress/egress times for the simulated monotransiting sample. For each simulated planet found to have a single detectable transit in the TESS primary mission we calculate $t_{in/eg}$ using equation \ref{eq:t_in/eg} and plot the distribution in figure \ref{fig:t_in/eg}.

% \begin{figure}[ht]
%     {\includegraphics[width=\columnwidth]{Figures/t_in_eg_plot.png}}
%     \caption{$t_{in/eg}$ values for all TESS primary mission monotransits.}
% \label{fig:t_in/eg}
% \end{figure}

% From this figure we can see that the number of monotransits falls off with increasing ingress/egress time. Since ingress/egress time scales with period these systems correspond to longer periods which are less plentiful in our simulation due to the occurrence rates of these type of systems used in this simulation \citep{2015ApJ...807...45D,2013ApJ...766...81F}. The result of this is that only 19 out of 540 monotransits have a value of $t_{in/eg}>98.6$\,minutes, meaning they are too long to fit our criteria for CHEOPS observations.

\subsection{CHEOPS}
\label{sec:CHEOPS results}

% Using the values of $N$ and $N_{transit}$ we find the efficiency and thus minimum required observing time per CHEOPS orbit for each monotransit. Their distribution is shown in figure \ref{fig:req_time}. This distribution excludes the 307 monotransits with $V<6$ or $V>12$.

% \begin{figure}[ht]
%     {\includegraphics[width=\columnwidth]{Figures2/req_time.png}}
%     \caption{Distribution of minimum required observing time per CHEOPS orbit for all TESS monotransits with $6\leq V\leq12$.}
% \label{fig:req_time}
% \end{figure}

% This plot seems is heavily biased towards short required times. This is a result of the ultra-high precision of CHEOPS. The number of in transit data points required to confirm at transit at $S/N\geq7.3$ is almost always a very small fraction of the full transit width, especially for longer period systems (such as monotransits) with correspondingly longer transit durations.

% Figure \ref{fig:coverage+monos} shows these monotransits plotted over the CHEOPS sky coverage map for a requirement of at least 19\,minutes of observations per orbit. We choose 19\,minutes as it gives an efficiency of $\sim$\,20\%, more than sufficient compared with the values required for most TESS monotransits.

Using the specified observation efficiency of 40\% as above we plot the 433 TESS primary mission monotransits over the corresponding CHEOPS sky coverage plot.

\begin{figure}
    {\includegraphics[width=\columnwidth]{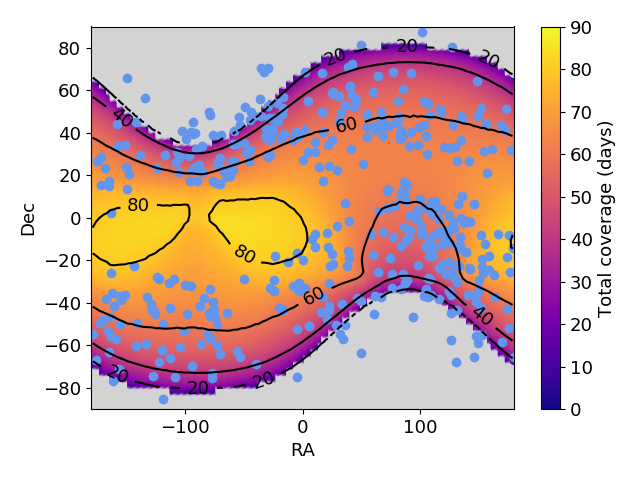}}
    \caption{TESS monotransits (blue) plotted over the CHEOPS sky coverage map for a minimum of 39\,minutes of observation per orbit ($\sim$\,40\% efficiency). The colour shows the total CHEOPS observing time in days across one year.}
\label{fig:coverage+monos}
\end{figure}

It can be seen from this plot that, based on the 39\,minutes sky coverage, only a small fraction of monotransits fall on uncovered sky. In fact, only 46 out of the 433 monotransits fall outside of the required CHEOPS coverage. This leaves 387 TESS primary mission monotransits observable by CHEOPS, allowing for our observing criteria ($\sim$\,89\%). An additional point that can be seen in this plot is the lack of TESS monotransits around the regions of sky which CHEOPS will observe for the longest. This is partially because these areas are along the ecliptic equator which TESS will not observe during its primary mission and partially because the shifted TESS northern sectors happen to coincide with this region of sky.

\begin{figure}
    {\includegraphics[width=\columnwidth]{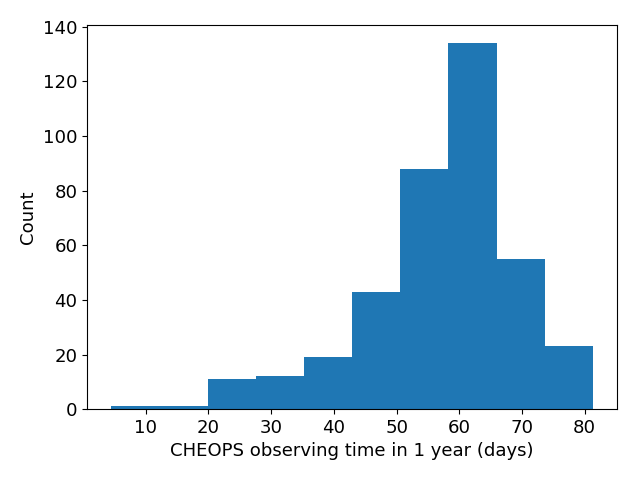}}
    \caption{Number of days per year for which each TESS primary mission monotransit could be observed by CHEOPS with at least 40\% efficiency. We show all 387 targets observable by CHEOPS.}
\label{fig:obs_time}
\end{figure}

The full distribution of CHEOPS observing time for the 387 observable TESS monotransits is shown in Figure \ref{fig:obs_time}.

The distribution of the observing time for the 387 observable targets is seen to peak around 60 days matching the distribution seen in Figure \ref{fig:coverage+monos}. This is a result of the majority of TESS monotransits being found in regions of sky with 1 TESS sector of observations, that is, between 6 and 30 degrees in latitude of the equator. These are the areas of sky which match up with the 60 day contour in Figure \ref{fig:coverage+monos}. The sharp drop off above 60 days is again the result of no monotransits being found along the ecliptic equator, combined with the placement of the shifted TESS sectors. The drop off towards less CHEOPS observation is a result of the decreasing number of monotransits that are discovered further from the ecliptic equator, where the sky has multiple TESS sectors of coverage leading to more multitransit detections. This corresponds to the regions where CHEOPS has less coverage as well.

\begin{figure}
    {\includegraphics[width=\columnwidth]{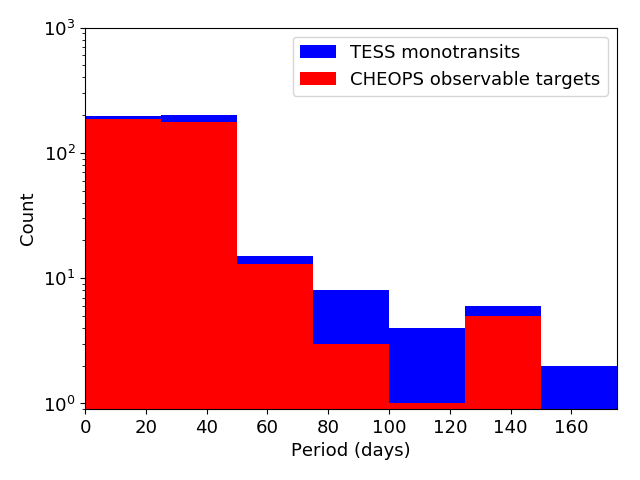}}
    \caption{Period distribution of monotransits. All TESS primary mission monotransits (433) are in blue and the subset that CHEOPS could observe based on our criteria (387) are in red.}
\label{fig:period_dist}
\end{figure}

An interesting aspect of this distribution to consider is which regions of the monotransit distribution can be recovered by CHEOPS. Figure \ref{fig:period_dist} show the period distribution of all 433 TESS primary mission monotransits as well as the subset which CHEOPS could observe.

From this plot we see that the TESS monotransits period distribution is similar to those seen in \cite{2018A&A...619A.175C} and \cite{2019A&A...631A..83C}. The distribution drops off with period due to the underlying occurrence rates used in this simulation, however we still find over 200 detections with $P\geq25$\, days and $\sim$\,30 detections at $P\geq50$\,days. The CHEOPS observable subset follows the TESS distribution shape closely but falls off faster with period. For the range $P<25$\,days CHEOPS can observe 96\% of TESS monotransits. This fraction is 88\% for $25\leq P<50$\,days and 87\% for $50\leq P<75$\,days. This then falls to an average of $\sim$\,50\% recovery for $P\geq 75$\,days. The drop off towards longer period systems is explained by the relation between TESS and CHEOPS sky coverage. Longer period monotransits are preferentially found in areas of the sky covered by multiple TESS sectors which are towards the ecliptic poles. However, these are also the areas where CHEOPS has the least coverage (see Figure \ref{fig:coverage}).

\begin{figure}
    {\includegraphics[width=\columnwidth]{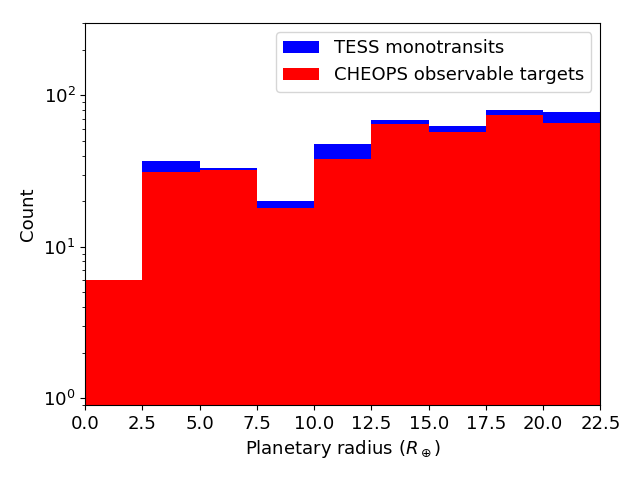}}
    \caption{Planetary radius distribution of monotransits. All TESS primary mission monotransits (433) are in blue and the subset that CHEOPS could observe based on our criteria (387) are in red.}
\label{fig:radius_dist}
\end{figure}

Figure \ref{fig:radius_dist} shows the corresponding planetary radius distribution.

As with period the CHEOPS population generally follows the TESS one closely. We see that the fraction of monotransits CHEOPS can recover is independent of planetary radius. This makes sense when comparing the ultra-high photometric precision of CHEOPS with that of TESS. Due to the increased precision the only factor preventing CHEOPS from picking up targets are their locations on this sky and unlike with period, there is no location dependence on planetary radius.

\subsection{CHEOPS timing}

There is still an additional effect to be considered however. Though we have shown which monotransits CHEOPS could observe in theory (those with non-zero sky coverage assuming a required efficiency of 40\%), it is not yet known if the times that they could be observed at will match up with additional transits. However, based on the periods of the monotransits and the sky coverage CHEOPS will achieve we can predict the number of monotransits that CHEOPS will be able to observe at the correct time to catch an additional transit. The precision of this forecast will be based on how well a period is known prior to using CHEOPS and is therefore an additional argument for photometric or spectroscopic data before CHEOPS is used.% {\bf TK: can we estimate the precision of this forecast? A high precision would be a strong argument in favor of observing a particular target with CHEOPS}.

We must first find the number of transits that will occur in a year, $n$, for each TESS monotransit. This is simply found by dividing a year by the orbital period. We then need the fraction of time CHEOPS could observe the target for within a year, $f$, found from the observing time (shown in Figure \ref{fig:obs_time}). The probability that CHEOPS will then be able to observe at least one additional transit, $F$, is given by the following equation

\begin{equation}
\label{eq:F}
    F = 1 - \left(1 - f\right)^{n}.
\end{equation}

Using this equation we find that CHEOPS would observe an additional transit for approximately 302 of the 387 monotransits that CHEOPS could observe.

\begin{figure}
    \begin{subfigure}[Orbital period
    \label{fig:F_period}]
    {\includegraphics[width=\columnwidth]{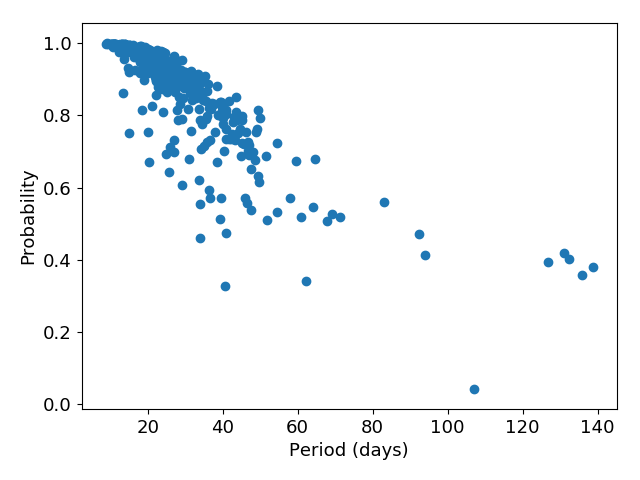}}
    \end{subfigure}
    \begin{subfigure}[Planetary radius
    \label{fig:F_radius}]
    {\includegraphics[width=\columnwidth]{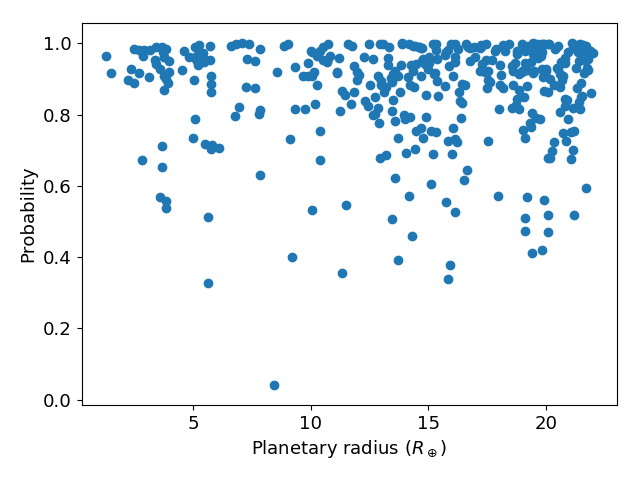}}
    \end{subfigure}
    \caption{Probability that CHEOPS will observe an additional transit of a TESS primary mission monotransit as a function of orbital period and planetary radius.}
\label{fig:F}
\end{figure}

Figures \ref{fig:F_period} and \ref{fig:F_radius} show how this probability changes as a function of orbital period and planetary radius respectively.

It can be seen from these plots that orbital period is the key parameter that determines whether another transit can be observed. We see a general decrease in $F$ with increasing period, as expected from equation \ref{eq:F}, and no real effect from changing radius. In terms of actual probabilities, for periods $\leq$\,50\,days the majority of monotransits have a greater than 80\% chance of having an additional transit observed with only a handful having probabilities less than 60\%. At longer periods the distribution becomes more sparse with many monotransits lying around 40-50\%.

\subsection{More conservative CHEOPS stability}

CHEOPS' orbit is foreseen to be very stable. Based on this stability it is valid to assume that CHEOPS is capable of detecting a transit from even a temporary drop of flux even if a whole ingress or egress is not seen or if the ingress or egress happens across multiple satellite orbits. However, CHEOPS is yet to return data and it is possible that its instrumental stability may be less than predicted. In this case a more rigorous criteria for a transit detection would be required and we make some comments to this effect here.

If the orbit-to-orbit stability is reduced we would need to be able to detect a transit from a single orbits worth of data alone. Therefore we now define a monotransit as observable by CHEOPS only if the observing time per orbit is sufficient to allow for the observation of a full ingress or egress (we assume this is sufficient for a transit detection). %{\bf TK: could you estimate the magnitude threshold given the current on-paper performances for this criterion?}
We also require that the per orbit observation time is sufficient that an ingress/egress cannot hide in the gap within a single CHEOPS period and be missed. Our minimum required observing time per orbit is then the larger of these two constraints. Based on this criteria there are then 2 cases where CHEOPS will not be able to sufficiently observe a target. First, ingress/egress duration is longer than a single CHEOPS period. Second, in the required coverage map the target coordinates fall in an uncovered region (marked grey in Figure \ref{fig:coverage}).% Assuming this is not the case we then interpolate on the chosen coverage map (noting this is now different for each monotransit) to calculate the total CHEOPS observing time for each monotransit target host.

Based on this interpolation we find that 176 TESS primary mission monotransits are observable for the required time per orbit by CHEOPS allowing for our more rigorous observing criteria. This outcome supports the fact that even should CHEOPS stability be less than expected a significant number of TESS primary mission monotransits could still be observed.

\section{Period estimation}
\label{sec:Period estimation}

As has been mentioned above CHEOPS will not be used as a blind follow-up for monotransits. In other words, CHEOPS will not stare at a monotransiting target waiting for a second transit to confirm its ephemeris, this would be an inefficient use of limited CHEOPS observing time. Therefore we need some estimate of period for these systems before CHEOPS will observe them. Predicting this period relies on obtaining additional observations of the systems in question, either photometric or spectroscopic.

As has been shown in \cite{2019A&A...631A..83C} $\sim$\,80\% of TESS primary mission monotransits will be seen to transit again during the TESS extended mission with $\sim$\,75\% transiting only once more. An additional transit during this mission means that we can now constrain the period into a discrete set of period aliases, which include the true period. In what follows we require the data from the TESS extended mission, therefore we limit ourselves to the southern ecliptic hemisphere as this will be re-observed earlier by TESS (July 2020 - June 2021) meaning that observing targets after the extended mission will still be feasible within the nominal CHEOPS mission lifetime (3.5 years). Additionally we focus on those systems that exhibit a single transit during each of the TESS primary and extended missions. This results in 132 planets.

Continuing our simulated TESS observations into the extended mission using the same procedure as \cite{2019A&A...631A..83C} we find that each system will have an average of 35 aliases when accounting for the TESS coverage and the separation between transits (for full details of this period alias simulation see Cooke et al. 2020, in prep.). 35 aliases is still too many for CHEOPS to reasonably target since most of these will be false and thus reveal no transit. Therefore we require additional observations to rule out these aliases. We simulate a suite of photometric and spectroscopic observations using the Next Generation Transit Survey at Paranal \citep[NGTS,][]{2018MNRAS.475.4476W} and CORALIE on the Euler 1.2\,m telescope \citep{2000A&A...354...99Q} respectively.

For photometric observations we simulate a stare campaign observing the target for a set amount of days using all night time hours as employed by \cite{2019MNRAS.tmp.2805G}. For spectroscopy we simulate one data point taken every 3 days for a set amount of time. After each night of observing we compare the simulated data and coverage with the set of period aliases based on the TESS transits. For photometry we look at each alias and compare its predicted transit times with the NGTS coverage to rule out aliases. For spectroscopy we fold our data on each alias and see if the phase curve is in line with the predicted shape based on the planet radius and spectrograph noise level (for more details of this period alias simulation see Cooke et al. 2020, in prep.).

\begin{figure}
    {\includegraphics[width=\columnwidth]{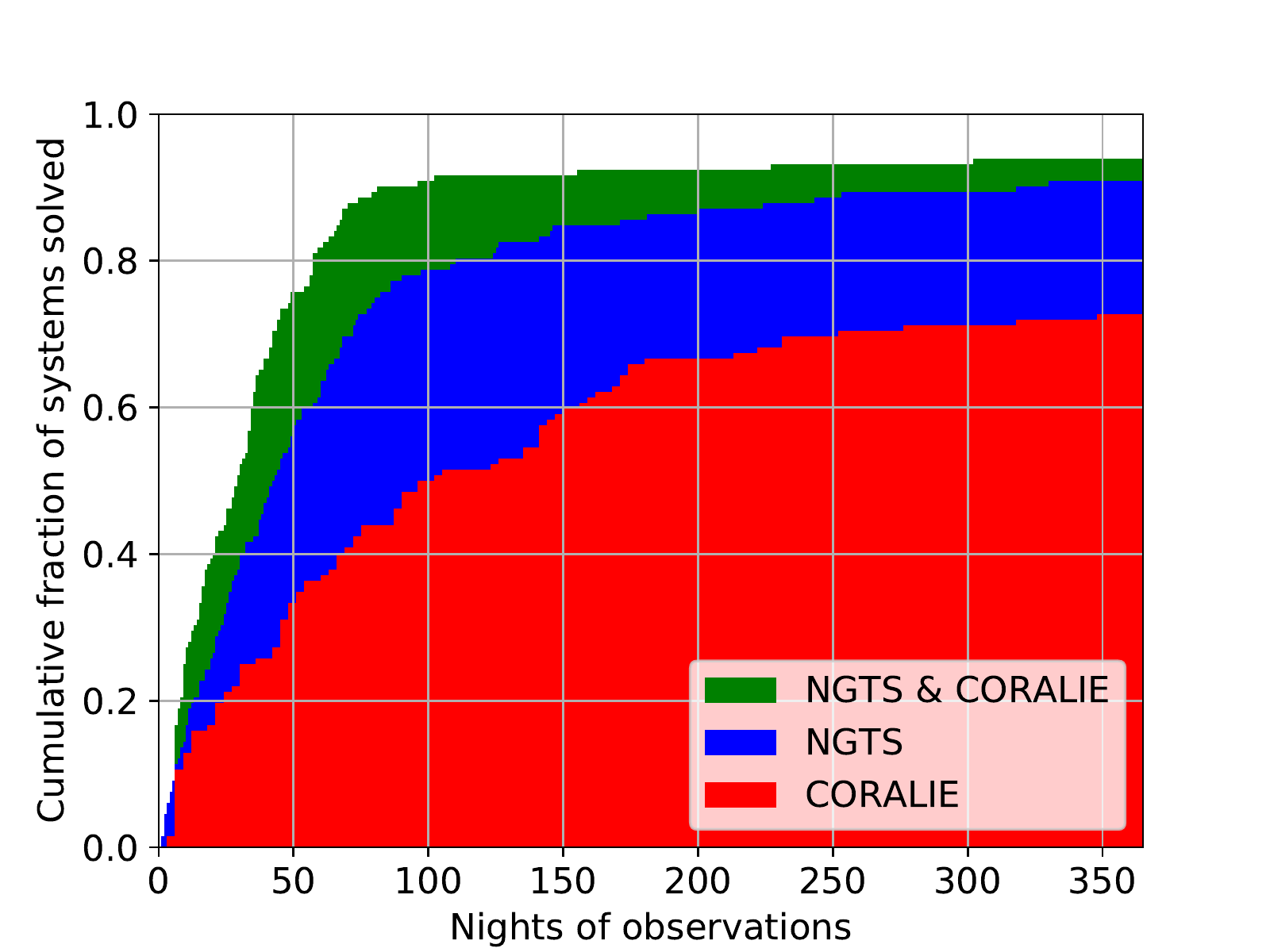}}
    \caption{Cumulative histogram of solved systems as a function of additional photometric or spectroscopic time. The distribution runs for one year with photometry being carried out every night by NGTS and spectroscopy consisting of one CORALIE point every three days. Only systems which have a solved period within 1 year are shown. Blue shows results using NGTS only, red using CORALIE only and green shows the combination of the two instruments.}
\label{fig:Solved_hist}
\end{figure}

The key result of this simulation for influencing CHEOPS follow-up is for what fraction of TESS primary mission monotransits can we determine the period before CHEOPS must observe it. Figure \ref{fig:Solved_hist} shows this distribution as a function of additional observing time.

We show that, depending on the amount of photometric and/or spectroscopic time used, up to 94\% of monotransits can have solved periods. Running the simulation for more time shows little change as the fraction of solved systems has plateaued. Using a combination of photometry and spectroscopy we show that 50\% of systems can have solved periods after $\sim$\,1\,month of additional observations per target. This means that, depending on the amount of photometric and spectroscopic time available, we will be able to target up to $\sim$\,125 TESS primary mission monotransits using CHEOPS, having already determined their periods. Extrapolating this into the northern ecliptic hemisphere (TESS will re-observe this hemisphere June 2021 - September 2022, still before the end of CHEOPS lifetime) gives the opportunity for CHEOPS to observe $\sim$\,250 TESS primary mission monotransits.

\section{Conclusions}
\label{sec:Conclusions}

We have shown that TESS will discover approximately 433 monotransits during its primary mission, including south and north ecliptic hemispheres. These systems are distributed across the sky as shown in Figure \ref{fig:monos} with period and planetary radius distributions as shown in Figures \ref{fig:period_dist} and \ref{fig:radius_dist}. For more details of the TESS monotransit population see \cite{2018A&A...619A.175C} and \cite{2019A&A...631A..83C}. We have then explored the feasibility of using CHEOPS to re-observe these monotransiting systems. Using a baseline observing efficiency of 40\% we have shown that 387 of the monotransits could theoretically be observed by CHEOPS ($\sim$\,89\%). We have also then shown how this observable distribution depends on orbital period and planetary radius finding that CHEOPS observations will slightly favour shorter periods but will not be affected by radius. Of these 387 monotransits we show that CHEOPS observations will be able to coincide with future transits for 302 ($\sim$\,78\%) of these systems.

For CHEOPS to realistically spend time following-up these systems we will require some knowledge of the period of the monotransit and we have outlined a method in which this may be obtained. We show that, by combining additional photometry from the TESS extended mission and NGTS with spectroscopy from CORALIE, up to 250 TESS primary mission monotransits will be able to have solved periods prior to CHEOPS' potential observations of them (for full details of this analysis see Cooke et al. 2020, in prep.). This would allow an efficient use of CHEOPS observing time to target transits and constrain the characteristics of these interesting long-period systems.

% We have determined this overlap based on the requirement that CHEOPS must be able to observe a star with sufficient coverage that it reaches $S/N\geq7.3$. Taking into account its limited magnitude range, we find that CHEOPS will be able to observe 242 of the 579 TESS monotransits. For CHEOPS to realistically spend time following-up these systems will require some knowledge of the period of the monotransit and we have outlined a method in which this may be obtained.

\section*{acknowledgements}

We thank the anonymous referee for their comments which helped to improve the focus and content of this paper. BFC acknowledges a departmental scholarship from the University of Warwick. The contribution at the University of Geneva by ML and TK, was carried out within the framework of the National Centre for Competence in Research "PlanetS" supported by the Swiss National Science Foundation (SNSF). ML also acknowledges support from the Austrian Research Promotion Agency (FFG) under project 859724 ``GRAPPA''. TK acknowledges the financial support of the SNSF. This project has received funding from the European Research Council (ERC) under the European Union’s Horizon 2020 research and innovation programme (project {\sc Four Aces}; grant agreement No 724427).

%%%%%%%%%%%%%%%%%%%%%%%%%%%%%%%%%%%%%%%%%%%%%%%%%%



\bibliographystyle{mnras}
\bibliography{TESS_CHEOPS.bib}

%\begin{thebibliography}{99}
%\end{thebibliography}

%%%%%%%%%%%%%%%%%%%%%%%%%%%%%%%%%%%%%%%%%%%%%%%%%%

%%%%%%%%%%%%%%%%% APPENDICES %%%%%%%%%%%%%%%%%%%%%

% \appendix

% \section{Some extra material}

% If you want to present additional material which would interrupt the flow of the main paper,
% it can be placed in an Appendix which appears after the list of references.

%%%%%%%%%%%%%%%%%%%%%%%%%%%%%%%%%%%%%%%%%%%%%%%%%%

% Don't change these lines
\bsp	% typesetting comment
\label{lastpage}
\end{document}